\documentclass[12pt]{article}
\usepackage{epsf}
\usepackage{epsfig}
\usepackage{amssymb,amsthm,amsmath}
\def\eqref#1{(\ref{#1})}

\newcommand{\calI}{{\cal{I}}}

\newlength{\myhspace}
\newtheorem{thm}{Theorem}[section]

\newtheorem{conj}[thm]{Conjecture}
\newcommand{\abs}[1]{|#1|}
\newcommand{\norm}[1]{\|#1\|}

\def\curl{\mathop{\rm curl}}

\def\grad{\mathop{\rm grad}}
\def\div{\mathop{\rm div}}
\title{What are the equations of motion of classical physics?}
\author{CARMEN CHICONE\footnotemark[2]}
\begin{document}
\maketitle
\renewcommand{\thefootnote}{\fnsymbol{footnote}}
\footnotetext[2]{Department of Mathematics, University of Missouri,
Columbia, MO 65211.\\
Email: carmen@chicone.math.missouri.edu}
\begin{abstract}
Action at a distance in Newtonian physics is replaced by finite
propagation speeds in classical physics, the physics defined by the field
theories of Maxwell and Einstein.
As a result, the differential equations of motion in Newtonian
physics are replaced in classical physics
by functional differential equations, where
the delay associated with the finite propagation speed 
(the speed of light)
is taken into account. Newtonian equations of motion, with
post-Newtonian corrections, are often used to approximate
the functional differential equations of motion. 
Some mathematical issues related
to the problem of extracting the ``correct'' approximate Newtonian
equations of motion are discussed.  
\end{abstract}


\section{Introduction}
This paper---an expanded version of a lecture presented in
the special session on
Applied Dynamical Systems of the CAIMS 2001 meeting in 
Victoria, BC---is 
an invitation to explore some of the mathematical issues related
to the foundations of post-Newtonian physics.

Let us recall that Newtonian forces 
(for example, the inverse square law for gravitation)
imply ``action at a distance.''  
This absurd, but outstandingly successful, premise of Newtonian theory
predicts that signals propagate instantaneously. In classical
physics, relativity theory postulates that signals propagate with
a velocity that does not exceed the velocity of light.
Thus, the forces of Newtonian physics must be replaced by force laws
that take into account the finite propagation 
speed of the classical fields---electromagnetic and gravitational---which 
determine the forces acting on a moving body. In turn,
the ordinary (and partial)
differential equations of Newtonian physics, which are derived from the
second law of motion
$d(mv)/dt=F$, 
must be replaced by 
corresponding functional differential
equations where the force $F$ is no longer a function of just
position, time, and velocity; rather, the classical force law
must take into account the time-delays
due to the finite propagation speed of the classical fields.

The functional differential equations of motion for classical field theory
are generally difficult, often impossible,
to express in a form that is amenable to analysis.  Thus, to obtain
useful dynamical predictions from realistic models, 
it is natural to
replace the functional differential equations of motion by 
approximations that are ordinary (or partial) differential equations
(cf. \cite{bel}).
Of course, Newton's equations are the premier choice for approximating
the true equations of motion. Indeed,  
due to the overwhelming success of Newtonian models in applied
mathematics---where in most cases characteristic velocities are
so low that relativistic 
effects are negligible---the dynamics of the true equations of
motion are often ignored. The purpose of this paper is to 
discuss some of the mathematical issues that must be addressed
to obtain a rigorous foundation for post-Newtonian dynamics, that is,
Newtonian dynamics with relativistic corrections taken into account.

I thank Bahram Mashhoon for suggesting several improvements of
the original draft of this paper.

\section{Newtonian versus relativistic physics}
\subsection{Newtonian physics}
Let us consider the motion of a body $p$ with mass $m$ that is influenced
by its \emph{Newtonian} gravitational attraction to a second body 
$P$ with mass $M$. 
In this case,
the equation of motion for the body $p$ is 
\begin{equation}\label{newtgrav}
m\ddot x=-\frac{GMm}{|x-y|^2}\frac{(x-y)}{|x-y|},
\end{equation}
where $x$ denotes the position of $p$ in space, $y$ denotes the position of 
$P$, and  $G$ is the universal gravitational constant. Note that
the force on the body $p$ changes continuously with the  position of $P$.
By viewing changes in the 
gravitational force detected at $p$ as a
signal produced by manipulating the position of $P$, the
(instantaneous) action at a distance of Newtonian gravity is 
seen to be equivalent to the infinite speed of propagation of gravity.

By coupling equation~\eqref{newtgrav} with the equation 
\[
M\ddot y=-\frac{GMm}{|x-y|^2}\frac{(y-x)}{|x-y|},
\]
which models the motion of $P$ as it is influenced by its gravitational
attraction to $p$, 
we obtain the prototypical model of Newtonian mechanics: a system
of ordinary differential equations for the  motion
of two bodies influenced by their mutual gravitational attraction.
The Newtonian model  for two charged particles  moving under the
influence of the Coulomb force is 
essentially the same; the
only change in the differential equations of motion 
is the replacement of the 
constant $-GMm$ by the product of the charges of the particles. 

There are Newtonian models for systems of bodies, 
for fluids, elastic media, etc. All of these models are
ordinary (or partial) differential equations. 

\subsection{Electrodynamics and gravitodynamics}
In classical physics, the fundamental electromagnetic and gravitational
forces are generally given by time-dependent fields that propagate at 
the speed of light. 

Maxwell's field equations determine the properties of the
electric and magnetic fields that influence the motion of charged
particles through the Lorentz force law. 
In fact, the motion
of a relativistic charged particle $p$ with mass $m$ and velocity $v$, 
which is
influenced by the electric field $E$ and magnetic field $B$
produced by a charged particle $P$,
is  given by 
\begin{equation}\label{eq:lfl}
\frac{d}{dt}\Big(\frac{m v}{(1-|v|^2/c^2)^{1/2}}\Big)
=q(E+\frac{1}{c}\,v\times B),
\end{equation}
where $c$ is the speed of light and $q$ is the charge on the particle $p$.

The electric field $E$ that affects $p$ at position $x$ at time $t$ 
is produced by $P$ at $y$ 
at the retarded time $t-\tau$ where 
\[
|x(t)-y(t-\tau)|=c\tau 
\]
(distance=speed$\times$time).
Hence, the electric field at $(x,t)$ is given by
\[
E(x,t)=f(t, x, \tau, y(t-\tau), \dot y(t-\tau),\ddot y(t-\tau)),
\]
where $f$ is some smooth function. The magnetic field
has a similar form. It follows that the Lorentz force 
acting in spacetime at $(x,t)$ is delayed by time $\tau$
(itself an implicitly defined function of space and time) as 
$p$ and $P$ move, and therefore the 
equations of motion are a coupled system of retarded functional
differential equations.

Explicit representations of $E$ and $B$ are obtained from
Maxwell's field equations
\[\begin{array}{ccccccc}
\div E&=&4\pi \rho,\qquad && \curl E&=&-\frac{1}{c}\frac{\partial B}{\partial t},\\
\div B&=&0,\qquad                  && c\curl B&=&4 \pi j
                                    +\frac{\partial E}{\partial t},
\end{array}
\]
where $\rho$ is the charge density, $j$ is the current density, and 
$c$ is the speed of light.  In addition, we have the 
conservation of charge: $\div j=-\partial\rho/\partial t$. 

A standard computation using vector calculus shows that
the fields $E$ and $B$  are given by
\begin{equation}\label{eq:phiA}
E=-\frac{1}{c}\frac{\partial A}{\partial t} -\grad\phi,\qquad \quad B=\curl A,
\end{equation}
where, once the Lorentz gauge condition 
\[\div A+\frac{1}{c}\frac{\partial\phi}{\partial t}=0\]
is imposed, 
the scalar potential $\phi$ and the vector potential $A$
satisfy the following wave equations with sources:
\begin{eqnarray*}
\frac{1}{c^2}\frac{\partial^2\phi}{\partial t^2}-\Delta \phi
   &=&4\pi \rho,\\
\frac{1}{c^2}\frac{\partial^2A}{\partial t^2}-\Delta A
   &=&\frac{4\pi }{c}j.
\end{eqnarray*}
Using classical potential theory, the wave equation, and variation
of parameters, the retarded-time potentials are given by
\begin{eqnarray*}
\phi(x,t)&=&
  \int \frac{\rho(y, t-|x-y|/c)}{|x-y|}\,dy,\\
A(x,t)&=&\frac{1}{c}\int \frac{j(y, t-|x-y|/c)}{|x-y|}\,dy,
\end{eqnarray*}
where the integrals are over all of space. The retarded-time 
electric and magnetic fields are then computed as in
display~\eqref{eq:phiA}. In practice, the current density is
given by $j=\rho v$. Hence, by the conservation of charge,
the charge density satisfies
the continuity equation 
\[\frac{\partial \rho}{\partial t}+\div (\rho v)=0.\] 

The electric and magnetic fields produced by an arbitrary
charge density are complicated. The potentials, however, can
be computed explicitly
for a point-charge moving in spacetime.
In this ideal case, the charge density is a Dirac-type measure 
associated with the moving point-charge, 
and the resulting retarded-time potentials---called the    
Li\'enard-Wiechert potentials---are given by
\begin{eqnarray*}
\phi(x,t)&=&\big(\, \frac{q}
    {|x-y|-\dot y\cdot(x-y)/c}\,\big)_{\,\mbox{ret}},\\
A(x,t)&=&\frac{1}{c} \phi(x,t)\, \dot y_{\,\mbox{ret}},
\end{eqnarray*}
where the subscript $\mbox{ret}$ indicates
that $y$ and $\dot y$ are evaluated at the retarded time $t-\tau$ and 
$\tau$ is given implicitly by the equation 
$c\tau=|x(t)-y(t-\tau)|$.
The associated electric and magnetic fields are computed using the
equations in display~\eqref{eq:phiA}, and the corresponding
equation of motion~\eqref{eq:lfl} for a point particle influenced by these
fields is relatively simple.
For example,
the motion of two charged particles confined to move on a line, with
only the Lorentz force taken into account, is modeled by 
the Driver-Travis system (see~\cite{driver,travis}) 
\begin{eqnarray*}
\frac{\ddot x_{i}}{(1-\dot x_i^2/c^2)^{3/2}}&=&
\frac{(-1)^i q_1q_2 (c+(-1)^i\dot x_j(t-\tau_{ij}))}
{c^2 m_i \tau^2_{ij} (c-(-1)^i\dot x_j(t-\tau_{ij}))},\\
c \tau_{ij}&=&|x_i-x_j(t-\tau_{ij})|,
\end{eqnarray*}
where  $(i,j)$  is in the set $\{(1,2),\, (2,1)\}$.

There is a similar equation for the gravitational two-body problem
(restricted to a line), but the equations are more complicated.
The basic reason is that Einstein's field equation
($R_{\mu\nu}-(1/2) g_{\mu\nu}R=8\pi\kappa T_{\mu\nu}$, 
where $R_{\mu\nu}$ is the
Ricci tensor, $R$ is the scalar curvature, $\kappa=G/c^4$, 
$T_{\mu\nu}$ is the stress-energy tensor,
$g_{\mu\nu}$ is the metric tensor, and the indices $\mu$ and
$\nu$ run over the integers from zero to four) is \emph{nonlinear}.
While the motion of a single test particle follows a geodesic in 
spacetime, the general relativistic two-body system seems to be  
too difficult to write down explicitly.  Models of this type are 
the subject of current research. For example, relativistic
effects, including radiation damping, are important
in the dynamics of binary neutron stars (see~\cite{cmr} and the references 
therein).
In modern physics, the theoretical study of gravitational dynamics
is generally more important than classical electrodynamics.
The reason is that quantum mechanics has superseded classical 
electrodynamics, but there is yet no
quantum theory of gravity.
At any rate, the true equations of motion  of the classical field
theories are functional differential equations.
\section{Post-Newtonian approximation}
The basic idea of post-Newtonian approximation, from 
a mathematical point of view, is the expansion of model equations
in powers of $1/c$. From a physical point of view, the idea is to consider
low velocity (compared with the speed
of light) weak field limits.
Note, for example, that  the relativistic form of  Newton's second law,
where the rate of change of the momentum is given by 
$\frac{d}{dt}(m v (1-\abs{v}^2/c^2)^{-1/2})$, reverts to 
Newton's law in the low-velocity limit. 
\subsection{Radiation damping}
\begin{quote}
Classical mechanics is a mathematically
consistent theory; it just doesn't
agree with experience. It is interesting, though, that 
the classical theory of electromagnetism is an
unsatisfactory theory
all by itself. There are difficulties associated with the \emph{ideas}
of Maxwell's theory which are not solved by and not directly associated with
quantum mechanics---R. Feynman~\cite[p. 28-1]{feynman}.
\end{quote}
According to Maxwell's field equations, a charged particle produces
electromagnetic fields as it moves. 
Since, in this case, a particle radiates energy,
it must slow down. This basic intuition has led to some thorny 
issues in physics.  To describe one of them briefly, let us consider
a sphere consisting of identical charged particles. 
As this body accelerates, the various charges
produce fields that affect the motion of the body through the Lorentz
force.

By considering motion along a line, Dirac reasoned that there are
retarded and advanced self-forces that have (on average over the
body)
the post-Newtonian expansions
\begin{eqnarray*}
F_{\mbox{ret}}
&=&\frac{\alpha q^2}{a c^2}\ddot x-\frac{2 q^2}{3 c^3}\stackrel{...}{x}
 +O(\frac{a}{c^4}),\\
F_{\mbox{adv}}
&=&\frac{\alpha q^2}{a c^2}\ddot x+\frac{2 q^2}{3 c^3}\stackrel{...}{x}
+O(\frac{a}{c^4}),
\end{eqnarray*}
where $x$ is the position of the centroid of the sphere,
$\alpha$ depends on the charge distribution ($\alpha=2/3$ for a round
sphere),  $q$ is the charge on one of the particles, and $a$ is the
radius of the sphere (see~\cite[Ch. 28]{feynman}). 
These forces are derived by expanding
the Li\'enard-Wiechert potentials (see~\cite[Ch. 9]{ll}).
 
The existence of advanced forces (preaccelerations)
would imply that the motion of a body is influenced by forces that 
are produced in its future. While this notion is surely problematic, it 
does lead to an interesting result. 

The ideal situation---Dirac's original motivation---is the theory
of the electron, an elementary particle that seems
to have no internal structure; it is supposed to be a point charge,
that is, a sphere with radius zero. One manifestation 
of the internal inconsistencies
mentioned in the quote by Feynman is  the  blowup of the coefficients of
the first
terms in these expansions as $a\to 0$. From Newton's second law, 
this coefficient,
$m_{\rm e}:=\alpha q^2/(a c^2)$,  represents a mass, called the
electromagnetic mass. Thus, the difficulty can be stated
as follows: the electromagnetic mass blows up as the radius of the
sphere shrinks to zero. 

The radius of the sphere is given by 
$a=\alpha(q^2/( m_{\rm e} c^2))$ and the number $q^2/( m_{\rm e} c^2)$
is called the classical electron radius ($\alpha$ is scaled out, because
this coefficient depends on the shape of the original charge distribution).
This value for the electron radius
does not agree with the value that can be computed directly from 
relativity theory, a fact that led to a crisis in classical physics 
that is still not completely resolved 
(see~\cite[Ch. 28]{feynman} for an extended discussion).

Dirac proposed a way to remove the apparent blowup of the 
electromagnetic mass. He
theorized that the true self-force (on a moving electron)
is half the difference of the
retarded and advanced forces; that is, 
\[F_{\mbox{self}}=\lim_{a\to 0}\frac{1}{2} (F_{\mbox{adv}}-F_{\mbox{ret}}).\]
Therefore, the self-force 
(the radiation reaction force) is
\[ F_{\mbox{self}}=\frac{2 q^2}{3 c^3}\stackrel{...}{x}.\]

Using Dirac's theory, a post-Newtonian model for the motion of an electron, 
confined to move on a line and with radiation reaction taken
into account, is given by the Abraham-Lorentz equation
\[
m\ddot x=\frac{2 q^2}{3 c^3}\stackrel{...}{x}+F,
\]
where $F$ is some external force.  In the presence of an electromagnetic
field, the equation of motion in space would be
\[
m\ddot x=q(E+\frac{1}{c}\,v\times B)+\frac{2 q^2}{3 c^3}\ddot v+F.
\] 
Since the particle radiates (produces
fields that carry energy) the self force should cause the particle
to lose energy and slow down. For this reason, the presence of 
the third-order time-derivative term
in the first differential equation is called radiation damping. 
Is this intuition correct; that is,  does the presence of this term cause
damping?

As a concrete example, consider
two identical (isolated) charged particles each with unit mass that 
are one unit apart at rest.
Imagine that these bodies are 
elastically connected by a force that can be modeled
by  Hooke's law with unit spring constant. 
Their relative motion, ignoring their mutual Coulomb and 
gravitational interactions,
is modeled by the differential equation
\[
\ddot x+x-1=\epsilon\stackrel{...}{x}, 
\]
where $\epsilon\sim c^{-3}$. A similar scenario for two bodies
with gravitational radiation damping taken into account,
leads to a \emph{nonlinear} oscillator of the form
\[
\epsilon z\frac{d^5 z^2}{dt^5}+\ddot z+z=1,
\]
where $\epsilon\sim c^{-5}$ (see~\cite{ckmr}).
In both cases, these post-Newtonian models are not Newtonian---for example,
the differential equations are not second-order.

Physical intuition suggests that the postulated two-body systems are
damped oscillators. This leaves open the mathematical question: Is damped 
oscillatory motion
predicted by these post-Newtonian models?

The electrodynamic model is linear. Its characteristic
equation
\[-\epsilon r^3+r^2+1=0\]
has the roots 
\[ 
i-\epsilon+O(\epsilon^2), \qquad -i-\epsilon+O(\epsilon^2), \qquad 
1/\epsilon+O(1);
\]
therefore, the first-order approximation of the general
homogeneous solution, which is given by  
\[
a e^{t/\epsilon}+b e^{-\epsilon t}\cos t+c  e^{-\epsilon t}\sin t,
\]
has a ``runaway'' mode. 
A similar result is true, but more difficult to prove, 
for the \emph{nonlinear} oscillator
\[
\epsilon z\frac{d^5 z^2}{dt^5}+\ddot z+z=1.
\]
Runaway solutions are clearly not physical. 
What do they represent? How should they be eliminated?
More precisely, we may ask: What is the correct \emph{Newtonian}
equation of motion with the radiation damping taken into
account?

There is an obvious \emph{mathematical} answer to our question
once  our post-Newtonian models are recognized as
singularly perturbed Newtonian equations. By the
definition of $\epsilon$ as the reciprocal of a power of the speed
of light, it is reasonable to assume that $\epsilon$ is a small parameter,
at least in the low-velocity regime. To recover the correct 
Newtonian model, we can apply Fenichel's geometric singular perturbation theory
(see~\cite{fen,fen2}).

For the electrodynamic oscillator, we have the equivalent (singularly
perturbed) first-order system given by 
\begin{eqnarray}\label{thesingsys}
\nonumber \dot x&=&y,\\
\nonumber \dot y&=&z,\\
\epsilon \dot z&=& z+x-1.
\end{eqnarray}
For the gravitodynamic oscillator, the appropriate
first-order system is given by
\begin{eqnarray*}
\dot z&=&u,\\
\dot u&=&v,\\
\epsilon^{1/3} \dot v&=&w,\\
\epsilon^{1/3} \dot w&=&x,\\
\epsilon^{1/3} \dot x&=&\frac{1-z-v}{2 z^2}-\epsilon^{1/3}\frac{5ux}{z}
 -\epsilon^{2/3}\frac{10vw}{z}.
\end{eqnarray*}
These systems are converted to regular perturbation problems by
rescaling time. 
The corresponding fast-time  electrodynamic first-order system
is obtained by the change of variables 
$s=t/\epsilon$ ($s=t/\epsilon^{1/3}$ for the gravitodynamic oscillator).
These fast-time systems are equivalent to their original slow-time
counterparts for $\epsilon\ne 0$. In effect, 
the rescaling of time produces a new family of systems, still
parametrized by $\epsilon$, but with a different
limit as $\epsilon\to 0$.  

The fast-time electrodynamic system is given by
\begin{eqnarray}\label{thesys}
\nonumber x'&=&\epsilon y,\\
\nonumber y'&=&\epsilon z,\\
z'&=& z+x-1;   
\end{eqnarray}
it is a regularly perturbed first-order system.
The corresponding unperturbed system ($\epsilon=0$) has a
two-dimensional invariant
manifold, $\{(x,y,z): z+x-1=0\}$, consisting entirely of rest points.
Moreover, this manifold is normally hyperbolic. 
In our  special
case,  where the invariant manifold consists entirely of rest points,
the requirement for normal hyperbolicity 
is that nearby trajectories are attracted to (or repelled from)  the
vicinity of the 
invariant manifold exponentially fast; or, in other words, the system matrix
of the linearized system at each rest point has a zero eigenvalue
with a two-dimensional spectral subspace (corresponding to the tangent
directions on the invariant manifold) and a nonzero eigenvalue with
a one-dimensional spectral subspace (corresponding to the normal
direction to the invariant manifold). 
For the unperturbed system~\eqref{thesys},
the nonzero eigenvalue $\lambda$ is the same at each rest point. In fact,
$\lambda=1$ and all nearby solutions are repelled
from the unperturbed invariant manifold at this exponential rate.

A fundamental result of Fenichel's
theory states that a normally hyperbolic invariant manifold persists.
Hence,
for each sufficiently small $\epsilon\ne 0$, the corresponding
system~\eqref{thesys} has
a two-dimensional normally hyperbolic invariant manifold (not necessarily
a linear subspace), called the slow-manifold. 
Moreover, the corresponding family of slow-manifolds 
depends smoothly on the parameter $\epsilon$ and each slow-manifold 
is invariant, normally hyperbolic, and repelling. 

The  original singularly perturbed slow-time family~\eqref{thesingsys} has
a corresponding family of normally hyperbolic invariant manifolds. Each
of them repels nearby solutions at an exponential rate. This accounts 
for the existence of runaway solutions. 

Using the invariance of
the perturbed slow-manifolds, it is easy to obtain (in local
coordinates) the equations of motion restricted to these
two-dimensional manifolds. The perturbed invariant manifold, 
for the fast-time system,
is the graph of a function of the form
\[z=h(x,y)=1-x+\epsilon h_1(x,y)+\epsilon^2 h_2(x,y)+O(\epsilon^3),\]
where the $h_i$ are functions to be determined. 
Moreover, the dynamical system on this invariant manifold
is given by
\[
x'=\epsilon y,\quad 
y'=\epsilon z=\epsilon(1-x+\epsilon h_1(x,y)+O(\epsilon^2)).
\]
For $\epsilon\ne 0$, the dynamical system on the corresponding slow-manifold
for the original system is thus given by
\[\dot x= y,\quad \dot y=1-x+\epsilon h_1(x,y)+O(\epsilon^2).\]

To solve for the functions $h_i$, we use the invariance of the perturbed 
manifold.
In effect, at each point on the manifold, each tangent vector
$X=(\epsilon y,\epsilon z, z+x-1)$ 
corresponding to the system of differential equations~\eqref{thesys},
is a linear combination
of the basis vectors $b_x=(1,0, h_x(x,y))$ and $b_y=(0,1, h_y(x,y))$ for
the corresponding tangent space on this manifold. The functions
$h_i$ are obtained
by equating coefficients of powers of $\epsilon$ in the vector identity
\[X=\alpha b_x+ \beta b_y,\] 
where $\alpha$ and $\beta$ are scalar variables (which are determined
by the first two components of this identity).
To first-order in $\epsilon$ and in the original slow-time, 
the equation of motion on the invariant
manifold is equivalent to the
(Newtonian) second-order system
\[ \ddot x+\epsilon \dot x+x=1,\]
a dynamical equation for a damped harmonic oscillator---just as
it should be. 

For the gravitodynamic two-body system, a similar analysis results in the
Newtonian dynamical equation
\[ \ddot z+32\epsilon \dot z z(z-\frac{15}{16})+z=1. \]
Again, this is a (nonlinear) 
under-damped oscillator, at least if  the initial separation between
the bodies  is near $z=1$. 
\subsection{Synthesis}
We have just seen that geometric singular perturbation theory
(in particular, reduction to the slow-manifold)  produces
Newtonian model equations with post-Newtonian corrections that
give physically reasonable dynamics; in particular, the runaway solutions
are eliminated. How can we justify using these models?
Note, for instance, that the slow-manifolds in our models are
\emph{unstable}; nearby solutions run away. In applied mathematics,
we usually justify approximations by their \emph{stability}. 
To validate the slow-manifold reductions, we must show that 
the resulting Newtonian model equations are ``stable'' with
respect to the 
dynamics of the original functional differential equations, 
the true equations of motion in classical physics. 

\begin{conj}     
In the low-velocity regime, a functional differential equation
of motion derived using the forces in classical field theory 
(the Lorentz force or the relativistic gravitational force)
has an inertial manifold $\cal I$
(a finite-dimensional, invariant, exponentially attracting, and
smooth manifold) such that the restriction of the motion
to this manifold is a Newtonian dynamical system.
The $N$th-order singular perturbation problem, 
obtained by (post-Newtonian) expansion
in powers of $1/c$ and truncation at order $N$, has an equivalent
first-order system with a normally hyperbolic slow-manifold
$\cal S$. The corresponding vector fields that generate the
dynamical systems on $\cal I$ and $\cal S$ agree to order $N-1$
in $1/c$.
\end{conj}

While the  post-Newtonian expansion (and truncation) results
in a system that might (and usually does) have 
runaway modes, these are simply artifacts of the singular
nature of the expansion. 
The long-term dynamics of the functional differential equations 
obtained by a direct application of classical field theory is 
given by a Newtonian equation on a finite-dimensional inertial
manifold, and the same dynamical system is obtained 
by appropriate reduction to a slow-manifold for the singular
high-order post-Newtonian equation. 
Runaway modes generally have no physical 
significance; equivalently, the slow-manifold is generally
\emph{not} an attractor.

The conjecture states a rigorous justification for the validity
of the post-Newtonian model that agrees with the dynamics on 
an inertial manifold which is supposed to exist in the low-velocity regime. 
But, what is a low velocity?
A deeper question addresses
this issue directly. For what range of parameter values does
the inertial manifold persist? 

In the mathematical analysis,
$1/c$, or (even better) a characteristic
velocity divided by $c$, is to be viewed as a small parameter.
If the conjecture is valid, then there is 
a lower bound for $\epsilon$ such that the corresponding 
functional differential
equation has an inertial manifold. A theorem meant
to validate the conjecture should include an estimate for this bound.

What accounts for the transition from the low velocity to the high velocity
regime? 
Answer: As a characteristic velocity increases, 
a bifurcation will occur that destroys
an inertial manifold and, therefore, the validity of post-Newtonian
approximation. How can such bifurcations be detected?

\section{Delay equations}
To test the conjecture stated in the last section,
let us replace the functional differential
equations of mathematical physics (with space-dependent delays)
with families of delay differential equations of the form 
\begin{equation}\label{bde} 
\dot x(t)=f(x(t),x(t-\tau)), \quad x\in {\bf R}^n, \quad \tau\in {\bf R},
\end{equation}
where the delay $\tau$ replaces the small parameter corresponding
to $1/c$ and $f$ is a smooth function.

The usual state space for the delay equation~\eqref{bde}
is $C([-\tau,0])$, the space of continuous functions that map
the interval $[-\tau,0]$ into ${\bf R}^n$. A basic result in the 
well-developed theory of delay equations (see~\cite{diek,driverb,els,hale})
states that the initial value problem consisting of equation~\eqref{bde}
and an initial function in $C([-\tau,0])$ has a unique solution
$t\mapsto x(t)$ for $t$ in some interval $[0, \beta)$, where
$\beta>0$ or $\beta=\infty$. For such a solution, the state of the system
at time $t>0$ is given by $x_t\in C([-\tau,0])$, where
$x_t(\theta)=x(t+\theta)$. This assignment defines a semi-flow in 
the state space given by $(T^t\psi)(\theta)=x(t+\theta)$, where
$t\mapsto x(t)$ is the solution with initial condition $\psi$.

\subsection{Inertial manifold reduction}
A heuristic argument indicates that the delay equation~\eqref{bde}
has an inertial manifold for small $\abs{\tau}$. Let us first
introduce a new variable
$t=s\tau$ so that $y(s)=x(s\tau)$ is a solution of the 
delay equation
\begin{equation}\label{tde}
\dot y(s)=\tau f(y(s),y(s-1))
\end{equation}
whenever $x$ is a solution of the delay equation~\eqref{bde}.
The state space for the family~\eqref{tde} is
$C([-1,0])$. Also, the
unperturbed system ($\tau=0$)
\[\dot y(s)=0\] 
generates the semi-flow given by
\[
T^t\psi=\left\{ \begin{array}{cc}
\psi(t+\theta),\quad &\mbox{$0\le t<1$  and $ t+\theta<0$},\\
\psi(0),\quad & {\rm otherwise.} \\
\end{array}\right.
\]
The $n$-dimensional submanifold
\[
{\calI}:=\{\psi\in C([-1,0]) :
\mbox{$\psi(\theta)\equiv a$ 
for some $a\in {\bf R}^n$}\}
\]
consists entirely of rest points and is normally hyperbolic. In fact, every solution
reaches $\calI$ in time 
$t=1$, a rate of normal contraction that is faster than
any exponential decay. Thus, it is reasonable to expect that
${\calI}$ persists as an inertial manifold for sufficiently
small $\abs{\tau}$.

We have not proved the persistence of the invariant manifold. 
Because the state space is infinite-dimensional, there are some delicate issues involved in proving 
the existence and smoothness of an inertial manifold. Recent
results (see, for example, \cite{blz}) on the persistence
of infinite-dimensional normally hyperbolic invariant
manifolds may apply. But, at the first level, a theorem in this direction
would state the existence of an inertial manifold
for sufficiently small values of the parameter. 
This would not be completely satisfactory for future 
applications to physics where
some estimate of the relevant parameter values would be required.
The foundation for a direct proof of the existence of an inertial
manifold, with explicit bounds but without a proof of the required smoothness,
is contained
in the work of Yu.\ A.\ Ryabov and R.\ D.\ Driver (see~\cite{driver68,driver84}).
The smoothness of the inertial manifold is proved in~\cite{chiconenew}. 

Having indicated that an inertial manifold exists, let us reduce
the dynamical system to the inertial manifold and thus obtain the
``Newtonian system'' corresponding to the delay equation. To do this,
suppose that $\xi$ is a coordinate on ${\bf R}^n$ and $y(t,\xi,\tau)$
is the flow on the inertial manifold; that is; $t\mapsto y(t,\xi,\tau)$
is the solution of the delay equation such that $y(0,\xi,\tau)=\xi$.
More precisely, 
$\psi(\theta)\equiv \xi$ is the initial condition for the solution
$y$ of the delay equation. 
With this notation,
we have that 
\[
\dot y(t,\xi,\tau)=f(y(t,\xi,\tau),y(t-\tau,\xi,\tau));
\]
hence, the vector field that generates the flow $Y$ on the inertial manifold
is given by 
\[X(\xi,\tau):=\dot y(0,\xi,\tau)=f(\xi,y(-\tau,\xi,\tau)).\]
Its expansion at $\tau=0$ is  
\begin{eqnarray}\label{Xiner}
\nonumber X(\xi,\tau)&=& f(\xi,\xi)-\tau D_2f(\xi,\xi)f(\xi,\xi)
+\frac{\tau^2}{2!}
\{D_2^2 f(\xi,\xi)(f(\xi,\xi),f(\xi,\xi))\\
&& {}+
D_2 f(\xi,\xi)(D_1 f(\xi,\xi)+3 D_2 f(\xi,\xi))f(\xi,\xi))\}+O(\tau^3),
\end{eqnarray}
where the operator $D_1$,  respectively $D_2$, denotes differentiation
with respect to the first, respectively the second, argument of $f$.

It is interesting to note that the presence of a small delay in 
a conservative system often 
results in damped long-term dynamics on an associated inertial manifold.
For example,  
the Duffing-type model equation
\[
\ddot x+\omega^2 x=-a x(t-\tau)+b x^3(t-\tau)
\]
with small delay $\tau$ in the restoring  force, 
reduces (by a formal computation to first-order in $\tau$) 
to the van der Pol-type model equation 
\[
\ddot x+\tau(3 b x^2-a)\dot x+(a+\omega^2) x-b x^3=0
\]
on its inertial manifold. This example illustrates a phenomenon that is
reminiscent of quantization: while most periodic solutions in 
one-parameter families of periodic
solutions in a conservative system disappear in the presence
of a small delay, some persist as limit cycles. Does this observation
have physical significance?

\subsection{The analog of post-Newtonian expansion}
For the delay equation~\eqref{bde}, the analog of post-Newtonian
expansion is the expansion of the function 
$\tau\mapsto f(x(t),x(t-\tau))$
at $\tau=0$, where the first few terms of the series are given by 
\begin{eqnarray*}
f(x(t),x(t-\tau))&=&f(x(t),x(t))
 - \tau D_2 f(x(t),x(t))\dot x(t)\\
&&{} + \frac{\tau^2}{2!}( D_2 f(x(t),x(t))\ddot x(t)
+  D_2^2 f(x(t),x(t))(\dot x(t),\dot x(t)))\\
&&{} + O(\tau^3).
\end{eqnarray*}
Truncation of the expansion at order $N$ in $\tau$ produces an
$N$th order ordinary differential equation of the form
\begin{equation}\label{nthor}
(-1)^N\frac{\tau^N}{N!} D_2 f(x,x)x^{(N)}=
F(x,\dot x,\ldots, x^{(N-1)}, \tau),
\end{equation}
the desired analog of a post-Newtonian
expansion in classical physics. Moreover, by setting
$\mu:=\tau^{1/(N-1)}$ and treating $\mu$ as a small parameter, 
we obtain a singularly perturbed
first-order system of the form  
\begin{eqnarray*}
\dot x&=&y_1,\\
\mu^N \dot y_1&=&y_2,\\
&\vdots&\\
\mu^N \dot y_{N-2}&=&y_{N-1},\\
\mu^N \dot y_{N-1}&=&F(x,y_1,\ldots, y_{N-1},\mu^{N-1}).
\end{eqnarray*}
Under the assumption that $D_2f(x,x)$ is invertible,
the geometric singular perturbation theory can be applied to 
prove that this system has an  
$n$-dimensional slow-manifold for sufficiently small $\tau\ne 0$.
\subsection{Synthesis} 
For delay equations, the family of flows restricted to the
family of slow-manifolds (parametrized by $\tau$)
is generated by the family of  vector fields $Y(\xi,\tau)$. 
A  result in~\cite{chiconenew}
states that this family of vector fields
agrees with $X(\xi,\tau)$, the family that generates
the flows on the inertial manifolds, to order $\tau^2$. 

For applications to physics, the result that $X$ and $Y$ agree to
order-two in $\tau$ is sufficient for many applications. Of course,
it is not difficult to show that these vector fields agree to 
order three, four, etc. On the other hand,
the complexity of the
terms that appear in the expansion increases rapidly with the order.
While the  equality of the low-order coefficients of $\tau$ in the 
two expansions can be proved by direct computation, there does not
seem to be an easy abstract
proof for the equality of all coefficients.  
Thus, the conjecture that $X$ and $Y$ agree to order $N-1$ (one order less
than the order of the ``post-Newtonian'' truncation) is open.
The hypothesis that $D_2f(x,x)$ is invertible should be sufficient
for the conjecture, at least in the case where $N$ is sufficiently large. 
Under the stronger
hypothesis that $D_2f(x,x)$ is infinitesimally hyperbolic (no
eigenvalues on the imaginary axis), no such restriction on $N$ should
be necessary. 
 
A apparent difficulty to be overcome in the proof of the conjecture for 
delay equations is encountered in the proof of the conjecture   
for the special case of linear delay equations 
(see~\cite{chiconenew}).
\begin{thm}\label{th:A}
If $A$ is invertible, $\abs{\tau}\norm{A}e<1$, and 
\[
\dot x(t)=Ax(t-\tau),
\]
then the family of vector fields that generates
the flow of this system on its 
family of inertial manifolds (parametrized
by $\tau$) agrees to order $N-1$ with the family of  
vector fields on the corresponding family of slow-manifolds for
the corresponding family of $N$th order systems~\eqref{nthor}. 
Moreover, the family of vector fields on the inertial manifolds is 
given by
\[
X(x,\tau)=\sum_{j=0}^{\infty}(-1)^j\frac{(1+j)^j}{(1+j)!}\tau^j A^{1+j} x.
\]
\end{thm}
The proof in~\cite{chiconenew} requires
the combinatorial identity 
\[
\sum_{i=0}^m \binom{m}{i}   
(\ell-1+i)^{i-1}(m+1-i)^{m-i-1}=\frac{\ell}{\ell-1} (m+\ell)^{m-1}.
\]
This nontrivial identity can be proved
using Abel's generalization of the binomial theorem; namely,
the identity
\[
\alpha\beta
\sum_{i=0}^m\binom{m}{i}
(\alpha+i)^{i-1}(\beta+m-i)^{m-i-1}
 =(\alpha+\beta)(\alpha+\beta+m)^{m-1}
\]
(see~\cite[p. 19]{gj}). It seems that a ``nonlinear'' replacement for this
combinatorial identity will be required 
to prove the analog of Theorem~\ref{th:A}
for the delay equation~\eqref{bde}.  



\begin{thebibliography}{99}
\bibitem{blz} Bates P W, K  Lu, and C Zeng (1999)
Persistence of overflowing manifolds for semiflows,
\emph{Comm. Pure. Appl. Math}. {\bf LII} 0983--1046.
\bibitem{bel} Bel L (1982) Spontaneous Predictivisation 
in: Relativistic Action at a Distance: Classical and
Quantum Aspects, ed. J. Llosa, \emph{Lect. Notes in Phys.}
{\bf 162} 21--49.
\bibitem{cmr} Chicone C, B Mashhoon, and D Retzloff (2000)
Sustained resonance: a binary system perturbed by gravitational
radiation, \emph{ J. of Phys. A: Math. Gen.} {\bf 33} 513--530.
\bibitem{ckmr} Chicone C,  S Kopeikin, B Mashhoon,  and D G Retzloff (2001)
Delay equations and radiation damping
\emph{Phys. Letters A} {\bf 285} 17--26.
\bibitem{chiconenew} Chicone C (2001)
Delay equations, inertial manifolds, and singular perturbation,
\emph{In preparation}.
\bibitem{diek} Diekmann O, S A van Gils, S M Verduyn Lunel,  and
H O Walther (1995)
\emph{Delay Equations: Functional-, Complex-, and Nonlinear Analysis}
(New York: Springer-Verlag).
\bibitem{driver68} Driver R D (1968)
On Ryabov's asymptotic characterization of the solutions of
quasi-linear differential equations with small delays \emph{SIAM Review}
{\bf 10}(3)  329--341.
\bibitem{driver84} Driver R D (1976)
Linear differential systems with small delays \emph{J. Diff. Eqs.}
{\bf 54} 73--86.
\bibitem{driverb} Driver R D (1977)
\emph{Ordinary and Delay Differential Equations}
(New York: Springer-Verlag).
\bibitem{driver} Driver R D (1984)
A neutral system with state-dependent delay \emph{J. Diff. Eqs.}
{\bf 21} 148--166.
\bibitem{els} El'sgol'ts L E (1966)
\emph{Introduction to the Theory of Differential Equations with
Deviating Arguments} Mclaughlin R J, Translator
(San Francisco: Holden-Day).
\bibitem{feynman} Feynman R P, R B Leighton, and M Sands (1964)
\emph{The Feynman Lectures on Physics}, Vol. 2 (Reading: Addison-Wesley).
\bibitem{gj} Gouldent I P and D M Jackson (1983)
\emph{Combinatorial Enumeration}
(New York: John Wiley \& Sons).
\bibitem{hale} Hale J K and S M Verduyn Lunel (1993)
\emph{Introduction to Functional Differential Equations}
(New York: Springer-Verlag).
\bibitem{ll} Landau L D and E M Lifshitz (1971)
\emph{The Classical Theory of Fields} (Oxford: Pergamon Press).
\bibitem{fen} Fenichel N (1971)
Persistence and smoothness of invariant manifolds for flows
{\em Indiana Univ. Math. J.}
{\bf 21} 193--226.
\bibitem{fen2}
Fenichel N (1979)
Geometric singular perturbation theory for ordinary differential equation
\emph{J.~Diff.~Eqs.} {\bf 31} 53--98.
\bibitem{travis} Travis S P (1975)
A one-dimensional two-body problem of classical electrodynamics
\emph{SIAM J. Appl. Math.} (3){\bf 28} 611--632.
\end{thebibliography}
\end{document}